\documentclass[%
 aip,
 prl,
 amsmath,
 amssymb,
 twocolumn,
 10pt,
]{revtex4-2}

\usepackage{graphicx}% Include figure files
\usepackage{dcolumn}% Align table columns on decimal point
\usepackage{bm}% bold math
\usepackage{xcolor}
\usepackage[utf8]{inputenc}
\usepackage[T1]{fontenc}
\usepackage{mathptmx}

\usepackage{siunitx}
\usepackage{lipsum}  
\usepackage{wrapfig}

\definecolor{burntorange}{rgb}{0.8, 0.33, 0.0}
\definecolor{qnamigreen}{HTML}{48C1AC}

\newcommand{\BNV}[1]{$B_\mathrm{NV}$}
\newcommand{\figone}[1]{Fig.~\ref{fig1}{#1}}
\newcommand{\figtwo}[1]{Fig.~\ref{fig2}{#1}}
\newcommand{\figthree}[1]{Fig.~\ref{fig3}{#1}}
\newcommand{\figfour}[1]{Fig.~\ref{fig4}{#1}}
\newcommand{\figfive}[1]{Fig.~\ref{fig5}{#1}}

\newcommand{\muH}[0]{$\mu_0 H$}
\newcommand{\muHc}[0]{$\mu_0 H_c$}
\newcommand{\muHs}[0]{$\mu_0 H_{\mathrm{switch}}$}

\usepackage[acronym]{glossaries}

\newacronym{mram}{MRAM}{Magnetic Random Access Memory}
\newacronym{sttmram}{STT-MRAM}{Spin-Transfer Torque-MRAM}
\newacronym{sotmram}{SOT-MRAM}{Spin-Orbit Torque-MRAM}
\newacronym{vcmamram}{VCMA-MRAM}{Voltage Controlled Magnetic Anisotropy-Magnetic Random Access Memory}
\newacronym{vcma}{VCMA}{Voltage Control of Magnetic Anisotropy}
\newacronym{stt}{STT}{Spin Transfer Torque}
\newacronym{fl}{FL}{Free Layer}
\newacronym{rl}{RL}{Reference Layer}
\newacronym{hl}{HL}{Hard Layer}
\newacronym{tb}{TB}{Tunnel Barrier}
\newacronym{mtj}{MTJ}{Magnetic Tunnel Junction}
\newacronym{ap}{AP}{Anti-Parallel}
\newacronym{p}{P}{Parallel}
\newacronym{cipt}{CIPT}{Current In-Plane Tunneling}
\newacronym{vsm}{VSM}{Vibrating Sample Magnetometer}
\newacronym{tmr}{TMR}{Tunneling Magnetoresistance Ratio}
\newacronym{ra}{RA}{Resistance Area Product}
\newacronym{dtco}{DTCO}{Design Technology Co-Optimisation}
\newacronym{pma}{PMA}{Perpendicular Magnetic Anisotropy}
\newacronym{cmos}{CMOS}{Complementary Metal Oxide Semiconductor}
\newacronym{Keff}{$K_{eff}$}{effective anisotropy constant}
\newacronym{saf}{SAF}{Synthetic Antiferromagnet}
\newacronym{nvm}{NVM}{Nitrogen Vacancy Magnetometery}
\newacronym{snvm}{SNVM}{Scanning Nitrogen Vacancy Magnetometery}
\newacronym{nv}{NV}{Nitrogen Vacancy}
\newacronym{moke}{MOKE}{Magneto-Optical Kerr Effect}
\newacronym{cafm}{cAFM}{Conductive Atomic Force Microscopy}
\newacronym{mfm}{MFM}{Magnetic Force Microscopy}

\begin{document}

\preprint{AIP/123-QED}

\title[Publication in pre-print]{A quantum sensing metrology for magnetic memories}
% Force line breaks with \\

\author{Vicent J. Borr\`as}
 \email{vicent.borras@qnami.ch}
 \affiliation{Qnami AG, Muttenz}

\author{Robert Carpenter}
% \email{robert.carpenter@imec.be}
\affiliation{Imec, Kapeldreef  75, 3001 Leuven}

\author{Liza Žaper}
 \affiliation{Qnami AG, Muttenz}
\affiliation{University of Basel, Departement of Physics}

\author{Siddharth Rao}
% \email{Siddharth.Rao@imec.be}
\affiliation{Imec, Kapeldreef  75, 3001 Leuven}

\author{Sebastien Couet}
% \email{sebastien.couet@imec.be}
\affiliation{Imec, Kapeldreef  75, 3001 Leuven}

\author{Mathieu Munsch}
% \email{mathieu.munsch@qnami.ch}
\affiliation{Qnami AG, Muttenz}

\author{Patrick Maletinsky}
% \email{patrick.maletinsky@unibas.ch}
\affiliation{University of Basel, Departement of Physics}

\author{Peter Rickhaus}
 \email{peter.rickhaus@qnami.ch}
\affiliation{Qnami AG, Muttenz}

\date{\today}

\begin{abstract}
Magnetic random access memory (MRAM) is a leading emergent memory technology that is poised to replace current non-volatile memory technologies such as eFlash. 
%However, due to the stochastic nature of the MRAM writing process into nanoscale magnetic layers, device-to-device variability is heavily affected by extrinsic sources and poses a major limitation to the scaling of MRAM technologies. 
However, the scaling of MRAM technologies is heavily affected by device-to-device variability rooted in the stochastic nature of the MRAM writing process into nanoscale magnetic layers.
Here, we introduce a non-contact metrology technique deploying Scanning NV Magnetometry (SNVM) to investigate MRAM performance at the individual bit level. 
We demonstrate magnetic reversal characterization in individual, $<60~$nm sized bits, to extract key magnetic properties, thermal stability, and switching statistics, and thereby gauge bit-to-bit uniformity. 
%Finally, we demonstrate that a single SNVM stray field map enables the extraction of the relevant magnetic properties. 
We showcase the performance of our method by benchmarking two distinct bit etching processes immediately after pattern formation.
Unlike previous methods, our approach unveils marked differences in switching behaviour of fully contacted MRAM devices stemming from these processes. 
%Contrary to existing characterization methods, our approach reveals key differences in the MRAM switching performance resulting from the two processes that manifest in different switching behaviour of fully contacted MRAM devices.
Our findings highlight the potential of nanoscale quantum sensing of MRAM devices for early-stage screening in the processing line, paving the way for future incorporation of this nanoscale characterization tool in the semiconductor industry.
%Finally, compared to alternative, macroscopic measurements, this technique is especially sensitive to, and capable of measuring, tail bits.
\end{abstract}

\maketitle

%\begin{quotation}
%\end{quotation}

\section{Introduction}

\begin{figure}[t!]%{r}{0.5\textwidth}
    \includegraphics[width=86mm]{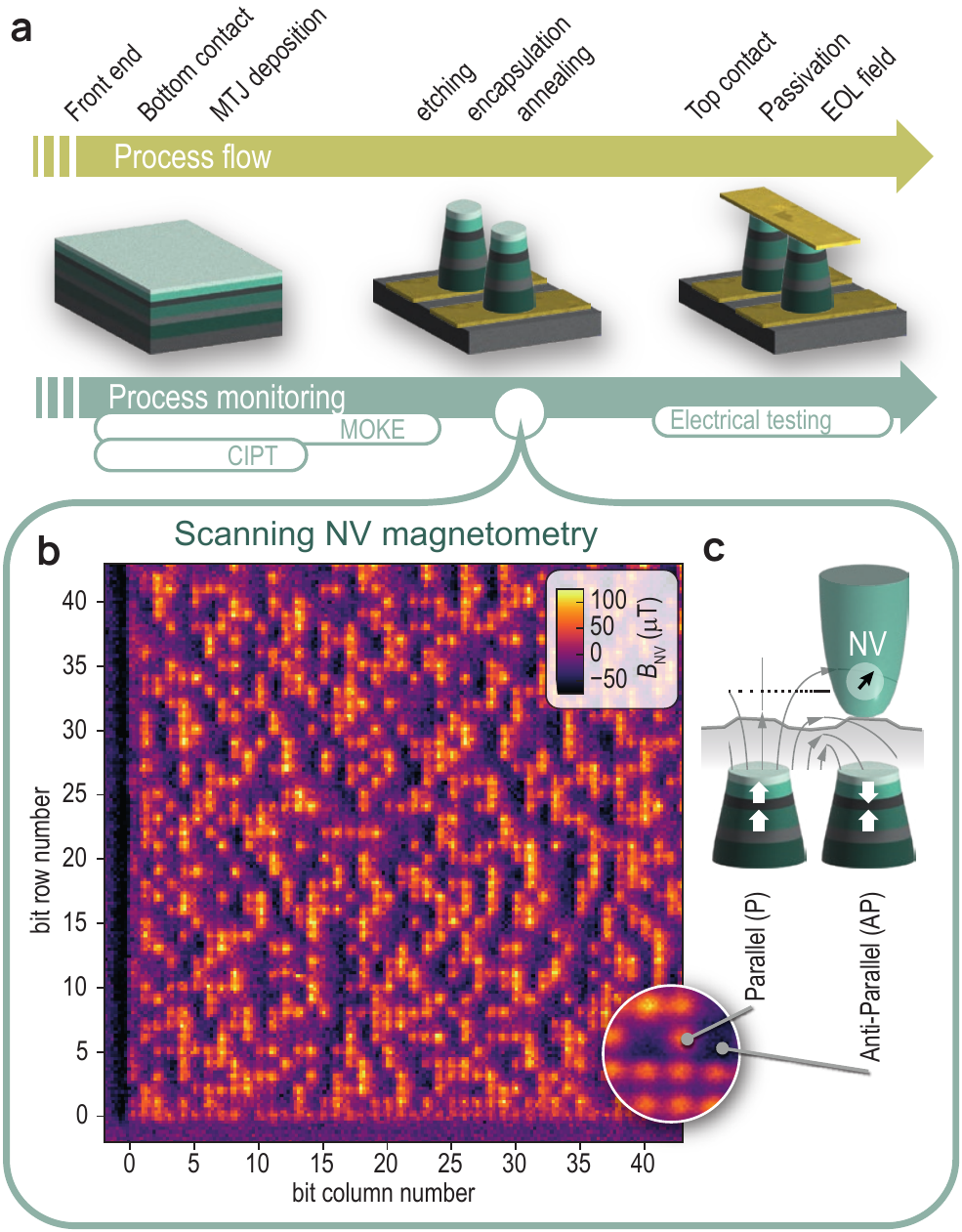}
    \caption{a) Process flow of STT-MRAM fabrication, including process monitoring. 
    b) SNVM map of 45x45 bits (10x$\SI{10}{um}$) after encapsulation. 
    Bits in the anti-parallel (AP) state appear dark, bits in the parallel (P) state appear bright.
    c) P and AP bit configurations generate distinct stray field patterns (gray lines). 
    The NV probe measures their projection onto the NV quantization axis (black arrow) at the flying distance of the NV probe.}
    \label{fig1}
\end{figure}

The ability to store information magnetically has been a major enabler of the digital revolution\,\cite{DataAge}. 
To improve storage density, and energy efficiency,  magnetic bits have become smaller and denser\,\cite{Dieny2020}, presenting an ever-increasing challenge for metrology tools -- one that is intensifying with the push towards exotic materials, such as 2-D and antiferromagnetic magnets, which have small surface moments\,\cite{Baltz2018,Du2023}. 
Emerging, ultrasensitive, nanoscale magnetic quantum sensors\,\cite{Rondin2014} offer a unique opportunity to address this challenge due to their highly competitive sensing characteristics. 
To demonstrate the advantage of such quantum meteorology in an industrially relevant context, and scale, \acrfull{sttmram} is an ideal candidate.
\acrshort{sttmram} is one of the most promising next generation, non-volatile memory architectures and one that is already in production\,\cite{Lee2019,Ji2020}. 
It is constructed around a \acrfull{mtj},  consisting of two magnetic layers, the \acrfull{fl} and the \acrfull{rl}, and a tunnel barrier.
The role of the \acrshort{fl} is to act as the storage layer, while the \acrshort{rl} generates a spin-polarised current. 
It is this spin-polarised current that, via \acrfull{stt}, can switch the \acrshort{fl} and read the bit state via the \acrfull{tmr} effect\,\ \cite{Miyazaki1995,Moodera1995}. %that occurs due to coherent tunnelling across the \acrshort{tb}\,\cite{Miyazaki1995,Moodera1995}.

Due to the interfacial nature of both the \acrshort{tmr} and \acrshort{stt} effects, uniformity is a significant challenge for \acrshort{sttmram}.
This is further exacerbated by the nanoscale dimensions of the magnetic layers, with typical target layer thicknesses, and bit sizes, of $<2~$ nm and $<60~$nm, respectively. 
In particular, the data retention, related to the energy barrier required to change the magnetic orientation of the \acrshort{fl}, is defined by the aniostropy and volume of the layer. Therefore, the energy to erase, or write, the \acrshort{fl} is volumetric and thus scales in quadrature with the diameter\,\cite{Peng2017}. 
This makes device variability especially sensitive to variations in the etch conditions.

While the device performance, and distributions are ultimately tested electrically\,\cite{OSullivan2018,Rao2021}, such characterization happens only at the end of the line, see \figone{a}. 
In-line measurements of bit-to-bit variability before electrical connection would enable early detection of faults, to monitor the processes at the most critical steps and to optimize these steps in shorter time frames. 
\acrfull{moke} magnetometry\,\cite{Kerr1877,McCord1995,Soldatov2017} and \acrfull{cipt}\,\cite{Worledge2003} are the most common methods for in-line process control.
However, these are only capable of measuring the as-deposited film properties and cannot provide the critical characterisation of individual bits after etching.
Recent works\,\cite{VanBeek2019,Choi2023} have shown that detailed properties of the magnetic film can be extracted using \acrshort{moke} magnetometry, however this method is only sensitive to the bulk of the distribution and not to single device performance. 
Alternative approaches to measuring single pillars are \acrfull{cafm} and \acrfull{mfm}\,\cite{Tryputen2016}. 
While \acrshort{cafm} enables electrical measurements after patterning, direct contact with a pillar is required which precludes straightforward use for in-line metrology. 
Conversely, \acrshort{mfm} can be used to map arrays, similar to the approach in this work. 
However, the resolution of \acrshort{mfm} is limited by the tip diameter which is typically $>30~$nm and the magnetic field sensitivity is typically lower. 
In addition, the technique is invasive and tip-to-tip variations can cause non-reproducible results, further limiting its application as an in-line measurement technique.

In this work, we show that quantum sensing based on \acrfull{snvm}\,\cite{Rondin2014, Maletinsky2012, RevModPhys.92.015004} can fill this key metrology gap. 
Using SNVM maps we estimate the retention, and characterize the bit-to-bit uniformity of two etch processes\,\cite{Rao2021b}. 
We find that while both processes result in significant bit-to-bit variations, the uniformity is improved with the optimized etch process. A single NV stray field map is sufficient to measure the improved uniformity.
Importantly, our results are consistent with prior electrical characterisation studies of the same processes\,\cite{Rao2021b}, which confirms the validity of our conclusion and the usefulness of \acrshort{snvm} as a future in-line characterisation tool.

%%%%%%%%%%%%%%%%%%%%%%%%%%%%%%%%%%
\section{Results} 
%%%%%%%%%%%%%%%%%%%%%

An example of an \acrshort{snvm} measurement on an STT-MRAM array is shown in \figone{b}, with details given in the Methods section. 
The image contains $45\times45$ bits and demonstrates the ability of \acrshort{snvm} to distinguish the logical state of individual bits. 
The image is obtained by continuously scanning the sample under an NV probe recording the projection of the local magnetic stray field  onto the NV quantization axis, see \figone{c}: 
Bits in the parallel (P) and antiparallel (AP) states (c.f. \figone{c}, right) yield comparatively high and low stray field magnitudes, leading to a bright and dark imaging contrast, respectively.

In order to demonstrate the key impact \acrshort{snvm} characterisation can have on \acrshort{sttmram} development, two wafers were fabricated in an industrial equivalent, $300~$mm wafer process. 
As one of the most critical steps in controlling \acrshort{sttmram} yield is the etch process, characterisation of this step was chosen as the main focus of this study. 
Thus two wafers were prepared using different etch processes, see \figtwo{a}. 
Process~1 consists of MTJ etching, oxidation, and encapsulation. 
In process~2, an additional etchback and gentle oxidation step is introduced\,\cite{Rao2021b}. 
Previous work has shown in electrically connected devices, that process~2 increases yield and decreases the bit-to-bit performance variation\,\cite{Rao2021b}.

\begin{figure}[t!]%{r}{0.5\textwidth}
    \includegraphics[width=86mm]{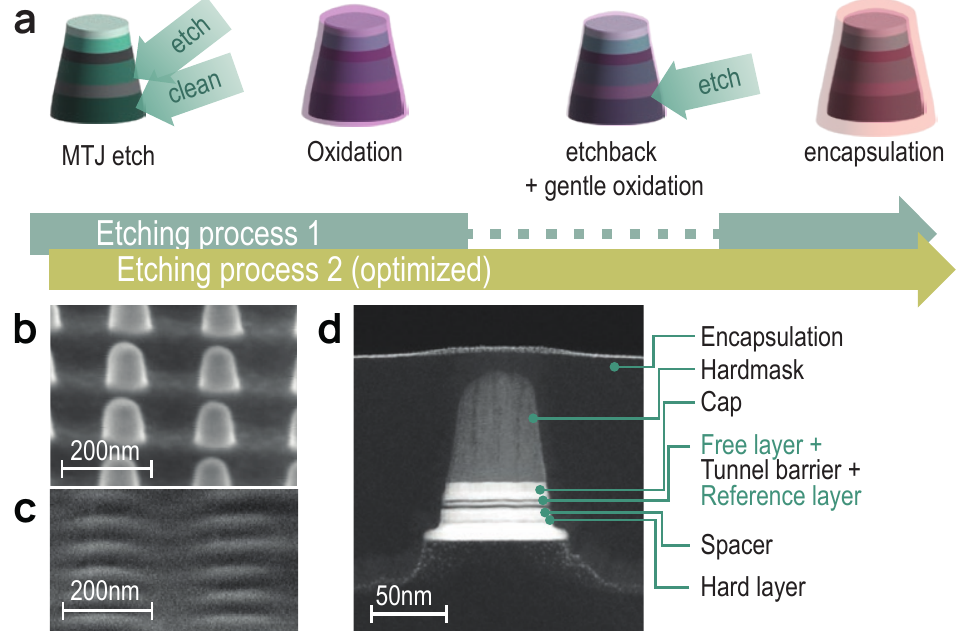}
    \caption{a) The quality of etching processes 1 and 2 are compared in this work. Etching process~2 features an additional etchback and gentle oxidation step after the first oxidation. 
    b) Scanning electron microscope images of the MRAM pillars before and c) after encapsulation.
    d) Transmission electron microscope image of an MRAM pillar.
    }
    \label{fig2}
\end{figure}

\begin{figure*}[ht!]
    \includegraphics[width=1\linewidth]{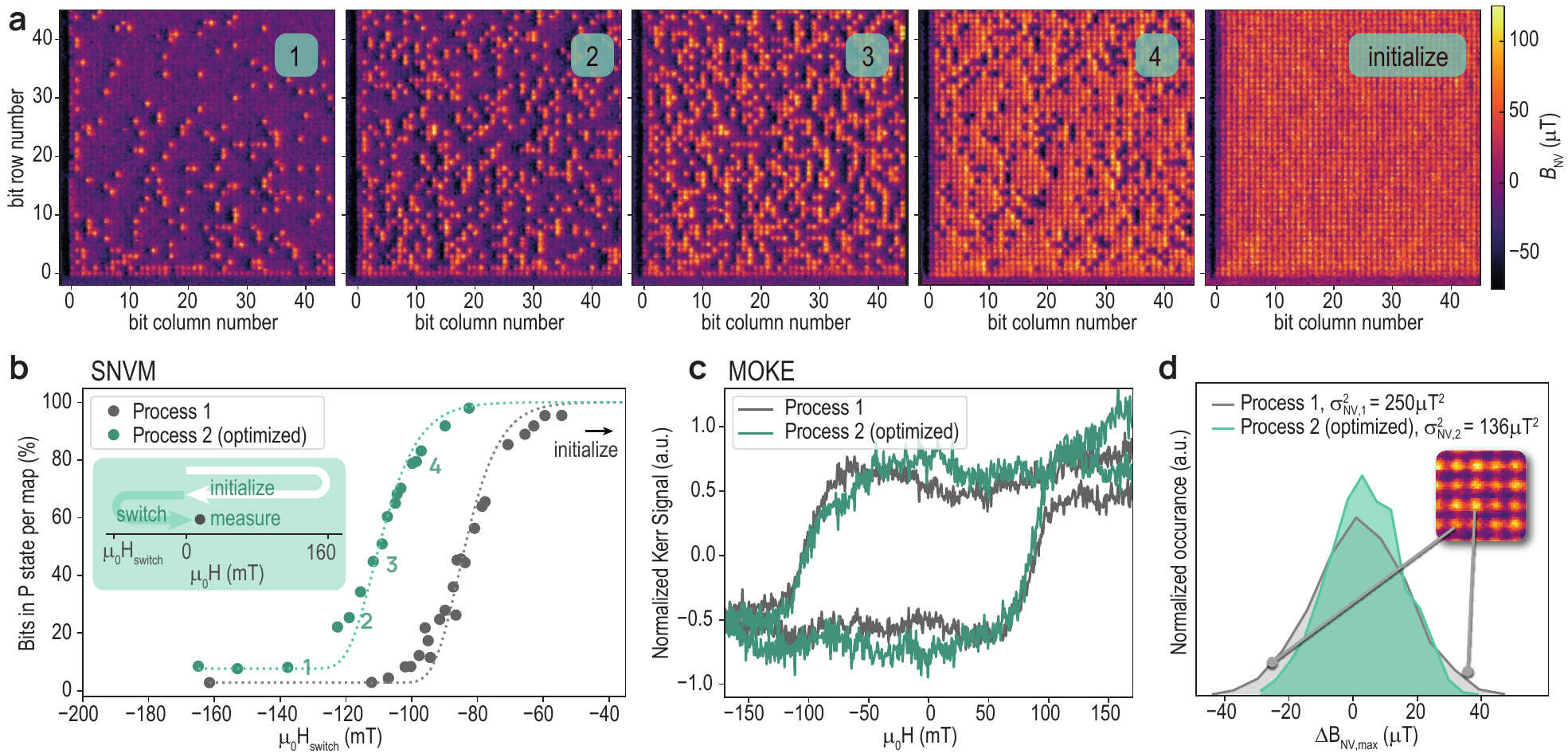}
    \caption{Quantitative determination of $\Delta$ and $H_k$. a) SNVM maps, obtained for different switching fields \muHs.
    b) For each map in a, the percentage of P bits is plotted versus \muHs\, which is fitted to obtain $H_k$ and $\Delta$, inset: measurement routine where the bits are first initialized at \muH=160 mT followed by a switching field \muHs. The measurements are performed at \muH=2 mT.
    c) Polar MOKE hysteresis loops of the FL measured on the same wafers. 
    d) Histogram of the maximum measured stray field per bit in P state,  shifted with respect to the median of the distribution, $\Delta B_{\mathrm{NV, max}}$.
    }
    \label{fig3}
\end{figure*}

The pillars were written with a $60~$nm diameter and a centre-to-centre separation of $200~$nm, but after  patterning, the physical diamater of the pillars is typically reduced to $\sim45~$ nm.  The \acrshort{mtj} tri-layer consists of a CoFeB \acrshort{fl}, an MgO tunnel barrier and a CoFeB \acrshort{rl}. 
SEM images before and after encapsulation are shown in \figtwo{b} and c, alongside a typical TEM image (\figtwo{d}) where the key layers are labeled. 
The samples were encapsulated with a SiN layer to limit physical damage due to sample preparation and oxidation from sample aging. 

We prepare and manipulate the bit states of the MRAM pillars by exposing them to an out-of-plane bias magnetic field, \muH, of varying strength. 
The resulting \acrshort{snvm} images in \figthree{a} are obtained for increasing values of \muH, using the measurement protocol illustrated in the  inset to \figthree{b}:
The bits are initialized in the P state by sweeping \muH\, to $160~$mT. 
We then switch a percentage of the bits to the AP state by sweeping to a negative field, \muHs, and then perform \acrshort{snvm} characterisation in a small bias field \muH$=2~$mT. 
Images in panels 1-4 in \figthree{a} are obtained with increasing values of \muHs\, where we observe a decreasing number of bits in the AP state. 
The image in panel 5 was measured directly after the initialization step, where all bits are aligned in the P state. 
Interestingly, we can initialize the array in a complete P state, but not in a complete AP state (\figthree{a}, panel~1), even when applying fields well above the average coercive field of the bits. 
This is seen in \figthree{c} where the preferred orientation is the P state. 
This asymmetry can be explained by the non-compensated stray field generated by the \acrshort{rl} and felt by the \acrshort{fl}. 
The asymmetry is particularly pronounced for bits at the edge of the bit array (row~1 and column~1, as well as row~2 and column~2), which indicates undesired magnetic properties of the bits close to the edge. Such difference may be explained by altered etching conditions at the edges of the arrays. Alternatively, the local stray fields may affect the neighbouring pillars. However, given the significant spacing of the pillars (the edge-to-edge spacing $>3\times$ the pillar diameter), such a cross-talk is unexpected\,\cite{Kim2007,Qui2009}. 
Even more so, the observation underlines the importance of local metrology of MRAM devices. 
In the following, we will exclude rows~1-4 and columns~1-4 from our data analysis, due to their unusual behavior.

To further analyze and quantify the bit switching process, we plot the percentage of parallel bits as a function of \muHs\, in \figthree{b}. 
The field at which $50\%$ of devices switch is found to be $\sim85~$mT and $\sim110~$mT for etch process~1 and 2 respectively. 
This is in excellent agreement with the coercive field, \muHc\, of the film (defined as half width of the \acrshort{moke} hysteresis loop), for which we find \muHc$\sim110~$mT by polar \acrshort{moke} magnetometry (\figthree{c}). 
These results provide another illustration of the advantage \acrshort{snvm} over traditional approaches, since \acrshort{snvm} clearly distinguishes the performance of the two etch processes, while \acrshort{moke} does not. 
The latter being limited by the large laser spot size employed, that leads to averaging over $\sim10^6$ bits and includes peripheral magnetic material that is in the field of the measurement. 
Therefore, while the polar \acrshort{moke} is well suited to measure the overall magnetic properties of thin films or bit ensembles, it is not sensitive to single devices or details of the performance distribution. 
Note that while micron-range spot sizes are available in \acrshort{moke} magnetometers, this is not sufficient to resolve individual bits and to the author's knowledge has never been applied to industrially relevant \acrshort{mram} devices. 

We proceed by fitting the \acrshort{snvm} data (dashed line, \figthree{b}) using the domain wall mediated reversal (DWMR) model\,\cite{7409773,mihajlovic2020thermal} in order to calculate the retention value ($\Delta$)
\begin{equation*}
    \Delta := \frac{E_b}{k_bT},
\end{equation*}
where $E_b$ is the energy barrier to switch between the P and AP states, $k_b$ is the Boltzmann constant and T is the temperature.  %To meet the industrial requirements the retention should be greater than 10~years, corresponding to $\Delta>80$\,\cite{}.
Fitting our \acrshort{snvm} data, we obtain values $\Delta=49$ and $\Delta=54$ for etch processes~1 and~2 respectively, indicating that the optimised process has a higher thermal-stability. 
In addition to the retention, the anisotropy fields ($H_k$) are calculated during the fitting and are found to be $\mu_0H_k= 440\pm2~$mT  for process~1 and $\mu_0H_k=533\pm2~$mT for process~2. 
For further details on the fitting, see Methods.

\begin{figure}[ht!]%{r}{0.5\textwidth}
    \includegraphics[width=86mm]{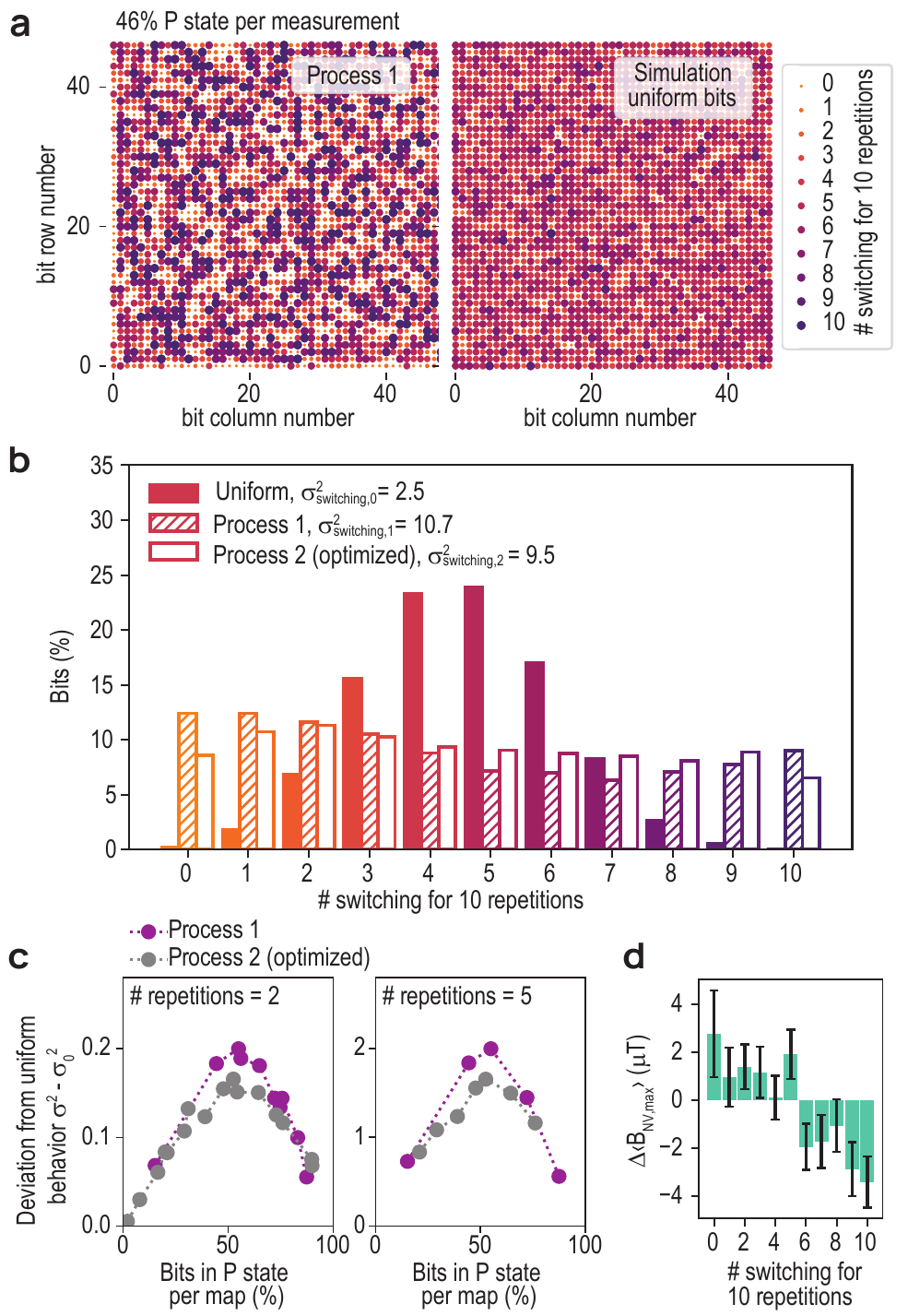}
    \caption{Array uniformity.
    a) The map on the left shows, by dot size and color, how often a bit has switched during 10 repetitive measurements. For each repetition, a map, as shown in \figthree{a} with  $46\%$ of bits in P state, is taken. The simulation illustrates how such a map would appear for perfectly uniform bits.
    b) Density histogram of the maps in a. The results are compared to a binomial distribution (purple) reflecting the expected outcome for perfectly uniform bits.
    c) Deviation from uniform behavior ($\sigma_\mathrm{switching, i}^2-\sigma_\mathrm{switching, 0}^2$) as a function of percentage of bits in P states for each individual measurement, we chose to show 2 and 5 repetitions. 
    d) Stray field in fully initialized map of Process~2, $\Delta B_{\mathrm{NV}, max}$ (see in \figthree{d}), as a function of \# switching of the corresponding bits. The stray field is averaged over all bits that show the same number of switching events. Errorbars indicate twice the standard error}
    \label{fig4}
\end{figure}

Due to the single-bit sensitivity of \acrshort{snvm}, we can furthermore extract valuable information by assessing the uniformity of the measured stray magnetic field.
\figthree{d} shows the distributions of the maximum measured  stray field $B_{\mathrm{NV, max}}$ per bit in the P-state for the two etch processes. 
It is clearly seen that process~2 has a narrower distribution 
with a stray-field variance $\sigma_\mathrm{B_{NV}}^2=136~\mu$T$^2$, as compared to that of process~1, where $\sigma_\mathrm{B_{NV}}^2=250~\mu$T$^2$. 
This indicates that process~2 has a higher bit-to-bit uniformity, in agreement with earlier findings\,\cite{Rao2021}.

In addition, we propose and demonstrate an alternative method to evaluate the homogeneity of the bit switching process across the array, which will confirm that $\sigma_\mathrm{B_{NV}}^2$ is indeed a valid qualifier for bit-to-bit performance uniformity. 
For this, we evaluate how often each bit successfully switches its state when exposing it several times to the same initialisation and switching conditions. 
We repeat initialization, switching and mapping 10~times by applying a value of \muHs\, for which almost half of the bits ($46\%$) remain in the P state for a single switching cycle. 
The statistics of switching events for a full bit-array is shown in \figfour{a} (left) for etch process~1, where the number of successful switches is color-coded in the dots representing the bits. 
In comparison to a simulation of bits obeying a binomial distribution (\figfour{a}, right), we note that a significant fraction of bits on the sample switch never (small yellow dots) or every time (large purple dots). 
This non-binomial behaviour is particularly pronounced at the edge of the array, as discussed earlier.

For a quantitative analysis, we show the corresponding density histogram in \figfour{b}. 
For Process~1, we observe that 12\% (9\%) of the bits switch zero (ten) times. 
Process~2 has a narrower distribution with 8\% (6\%), of zero (ten) switching events. 
For comparison, in an array following a binomial distribution, only 0.2\% (0.04\%) of bits would switch zero (ten) times. 
To quantify the non-uniformity, we extracted the variance of the distributions, and obtain $\sigma_\mathrm{switching, 0}^2=2.5$ for the binomial distribution and $\sigma_\mathrm{switching, 1}^2=10.7$ and $\sigma_\mathrm{switching, 2}^2=9.3$ for Processes 1 and 2 respectively. 
Although the deviation from the binomial distribution is large, there is clearly a more uniform behavior of the optimized Process~2 as compared to Process~1.

To confirm this observation, we repeat this experiment by changing \muHs\, to obtain different overall switching probabilities. 
In \figfour{c}, we show the deviation from the uniform behavior, $\sigma_\mathrm{switching, i}^2-\sigma_\mathrm{switching, 0}^2$ (with $i\in{1,2}$), as a function of the average percentage of bits in the P states for 2 and 5 repetitions. 
The graphs show a consistently high deviation, which peaks at a switching probability $\sim50\%$, which therefore marks the condition where the difference between the two processes becomes most evident. 
We note that, Process~1 consistently shows a stronger deviation from binomial behavior than the optimized Process~2, as is already apparent after two repetitions. The deviation from binomial switching is due to significant bit-to-bit variations of the relevant magnetic properties (magnetization, volume, shape, anisotropy) which is less pronounced for the optimized Process~2.

Importantly, the conclusion of these switching statistics agree well with the pattern observed in the stray field distribution (see \figthree{d}). 
For Process~2, the comparison of the average stray field measured from each bit to its switching behaviour (\figfour{d}), shows that those bits that are more (less) likely to switch produce, on average, a higher (smaller) magnetic stray field. 
This indicates that stray field distributions obtained with sensitive magnetic field measurements on initialized arrays can be used to retrieve information on the uniformity of the switching behavior.

\begin{figure*}[ht!]
    \includegraphics[width=1\textwidth]{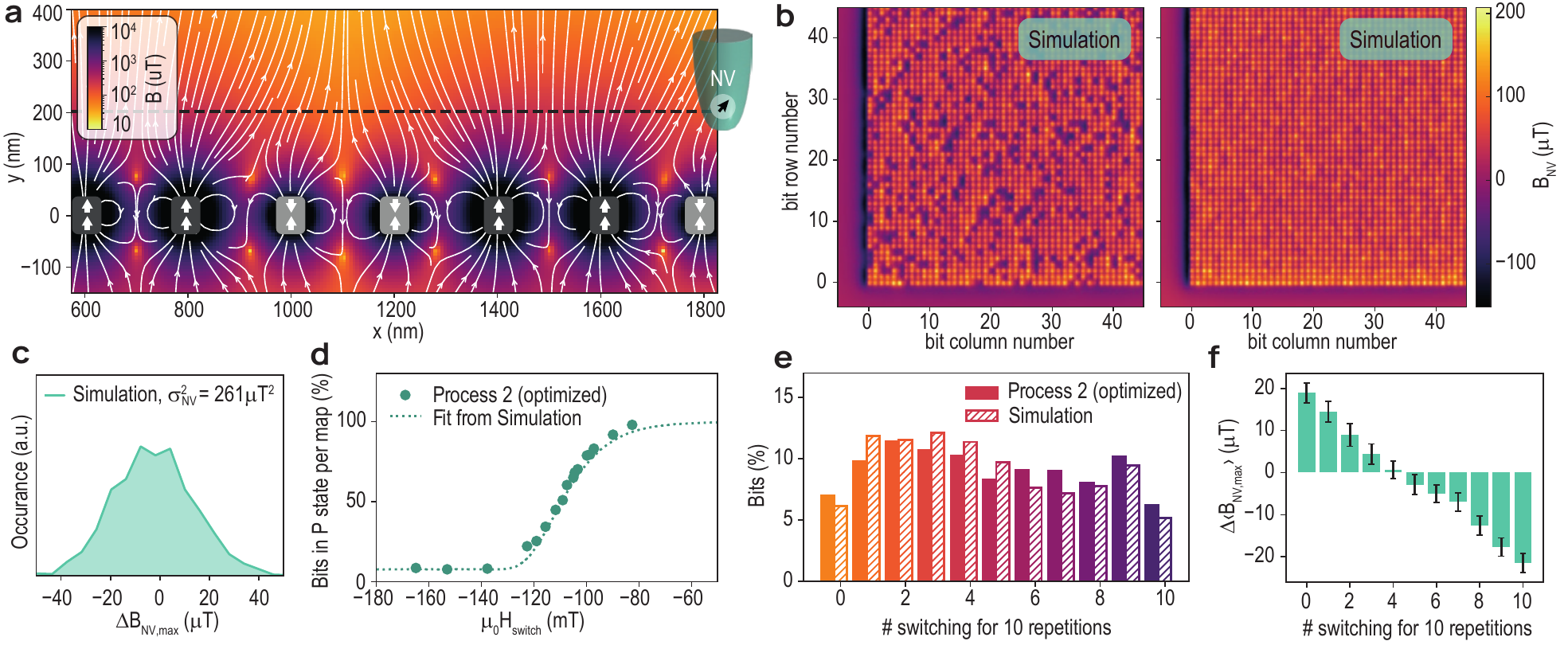}
    \caption{Magnetization distribution.
    a) Schema of the magnetic field obtained from a simulation for 7 pillars in different states (P or AP). 
    The dashed horizontal line  corresponds to the NV flying distance.
    b) NV maps obtained by simulations for 80\% and 100\% of the bits in the P state.
    c) Magnetic field distributions of maps in b.
    d) Percentage of P bits versus \muHs\, for process~2 (dots) and for the simulation (dotted line)
    e) Density histogram obtained after repeating the measurement 10 times at 50\% of switching probability.
    f) Stray field in fully initialized simulated maps, $\Delta B_{\mathrm{NV}, max}$, as a function of \# switching of the corresponding bits. 
    The stray field is averaged over all bits that show the same number of switching events. 
    Errorbars indicate twice the standard error}
    \label{fig5}
\end{figure*}

Finally, in order to understand the variations of the stray fields in terms of pillar properties, such as magnetization and critical dimension, we present simulations using the Magpylib library\,\cite{magpylib2020}. 
Here, the devices are modelled as cylindrical layers of different thicknesses and saturation magnetizations  (see \figfive{a}). 
The values for this simulation were taken from measurements made on non-patterned samples (data not shown, for further details, see Methods). 
The magnetic stray field is then calculated for a large array of pillars and projected onto the NV quantization axis at a height corresponding to the NV flying distance. 
This results in simulated maps that compare well to the SNVM measurements. 
\figfive{b} shows an example of a simulation in which we considered a normal distribution of the magnetization in the FL with a standard deviation of $0.235~$MA/m, a normal distribution in the critical dimension with $0.8~$nm of standard deviation and normal distributions in the magnetization directions with standard deviations of $10^\circ$ and $5^\circ$ for the RL and FL respectively. 
These values are taken according to recent work measured on similar samples to investigate the dispersion of easy-axes orientations\,\cite{frostswitching}. 
Taking these into account, a stray field histogram is generated \figfive{c}, which we find to be in good agreement with the one obtained from  \acrshort{snvm} data. 
Furthermore, assuming these distributions, the switching probability and the density histogram measured for etch process~2 are well reproduced (\figfive{d and e} respectively). 
Finally, in \figfive{f}, we plot the mean magnetic stray field as a function of the number of switching events, in order to reproduce \figfour{d}. 
Despite a difference in absolute value between experiment and simulation, the same trend is observed. 
Our simulation suggests that the non-binomial behaviour is related to the magnetic and structural variations among the pillars, such as the saturation magnetization, the magnetisation angle, and the pillar diameter. 
These variations are directly related to the key metrics for \acrshort{mram} device health, such as the thermal stability (see Methods) and read/write currents. 
This being measurable at the stage of the pillar etch is a significant step towards a rapid, early, detection of \acrshort{mram} process failures.

\section{CONCLUSION}
In conclusion, we demonstrate that SNVM is a powerful technique for the measurement and characterisation of \acrshort{mram} device properties early in the manufacturing process, directly after the devices etching step. 
Furthermore, high-resolution imaging of \acrshort{mram} pillars of industrially relevant diameters is achieved, even with encapsulation layers present. 
Our data are obtained with single-pillar resolution, a capability which is not available to present-day in/off-line metrology techniques. 
The quality of our \acrshort{snvm} images enables a quantitative analysis of the bit uniformity through a single stray-field map or a measurement of the switching behaviour. The latter can then be used as a method to extract the thermal stability with a sensitivity at the single-bit level. 
Finally, the flexibility of this technique is such that measurements normally limited to chip level could be enabled at an early stage. 
For example, by incorporating localized heating or perturbation fields, the probability of data loss in different application environments can be explored. 
These combined assets show the strength of nanoscale quantum metrology for addressing the magnetic ordering of patterned, magnetic, systems that are now in industry-scale production. 
Furthermore, the ability to sense even smaller magnetic fields such as those emerging from unsaturated spins in antiferromagnets\cite{Kosub2017, Hedrich2021} puts quantum sensing and the hereby demonstrated methodology  in a unique position for meteorology on a wide range of future magnetic memory architectures.

\section{METHODS}
\subsection{Fabrication method} 

 The deposition of the \acrshort{mtj} films was carried out by magnetron sputtering (Canon-Avelva EC7800) on 300 mm, thermally oxidized, Si wafers. 
 The thicknesses of all the layers were controlled through calibration of the deposition rates which, depending on the deposition conditions, are in the range of $0.1~$\AA/s to $1~$\AA/s. To simulate back end of line thermal budget conditions, the films were post-annealed at $400~^\circ C$ in a vacuum environment for 30 min (TEL-MS2 MRT5000). 
 These films were then patterned into large arrays of pillars with a design (final) diameter of $60~$nm ($45~$nm), with a centre to centre separation of $200~$nm. 
 The arrays were printed using immersion lithography, followed by patterning via ion beam etching. 
 The different etching approaches involved careful control of the side-wall etching and oxidation. 
 For a comparison of the two processes, please see previous work published by the imec team\,\cite{Rao2021b}.

\subsection{Scanning NV magnetometry} 
Our \acrshort{snvm} measurements were carried out on dies diced from the $300~$mm wafer using a diamond saw with an accuracy of $45~\mu$m. 
SNVM maps were recorded using the Qnami ProteusQ microscope.
 The SNVM technique is based on a \acrfull{nv}  which is implanted very close to the apex of a diamond pillar\cite{Maletinsky2012}. 
 The \acrshort{nv} center is highly sensitive to external magnetic fields\cite{RevModPhys.92.015004}. 
 While scanning the \acrshort{snvm} pillar over an MRAM array, the local magnetic stray field pattern is recorded. 
 This stray field is taken at a distance of $\approx\SI{150}{nm}$ to the free layer. The distance includes the cap, the hardmask an encapsulation and additionally the flying distance of the NV center on top of the surface ($\approx\SI{50}{nm}$). 
 The measured signal corresponds to the stray field vector at this distance, projected onto the NV-axis, which is tilted by $\phi=54.5^\circ$ with respect to the z-axis and has an angle in the x-y plane of $\theta\approx0^\circ$ with respect to the x-axis. 
 This projected magnetic field leads to a Zeeman splitting of the NV $\left|m_s\pm 1\right>$ state. 
 We follow the position of the $\left|m_s=- 1\right>$ state by recording the NV fluorescence versus applied microwave frequency in the $2.8~$GHz range (optically detected magnetic resonance, ODMR\cite{Maletinsky2012}). 

\subsection{Data analysis and simulations}
To obtain the bits array from the magnetic field map (e.g. \figone{b}) we prepared a script that finds the optimized grid in which the pillars are centered inside the grid squares. 
Then, depending on the mean value of the magnetic field each bit is classified as either 0 or 1.

To fit the data of \figthree{b} and obtain the $\Delta$ values we used the well known Néel-Brown relaxation model

\begin{equation}
    \label{eq1}
    P(H_{switch}) = 1-exp\left[\frac{-f_0H_{switch}}{R} exp \left[-\Delta (H_{switch})\right]\right],
\end{equation}
where the $f_0$ is the attempt frequency ($1~$GHz),  $R$ is the field sweep rate ($5~$mT/s) and $\Delta(H_{switch})$ is calculated assuming the domain wall mediated reversal (DWMR) model\,\cite{7409773,mihajlovic2020thermal} as 
\begin{equation}
\label{eq2}
    \Delta(H_{switch}) = \frac{1}{k_BT}\left[\sigma_{dw}l_{dw}\tau - \mu_0 H_{switch}M_s \tau \left[A_{+\delta_w}+A_{-\delta_w}\right]\right],
\end{equation}
where $k_B$ is the Boltzmann constant, $T$ is the temperature (298 K),  $\sigma_{dw}$ is the domain wall energy density, $l_{dw}$ is the domain wall length, $\tau$ and $M_s$ are the thickness ($1.2~$nm) and the magnetization ($1.178~$MA/m) of the free layer respectively, and $A_{-\delta_w/2}$ and $A_{+\delta_w/2}$ are the reversed and unreversed areas excluding the domain wall width $\delta_W$, which can be estimated with the following relation\cite{mihajlovic2020thermal}:
\begin{equation}
    \delta_{W} = 2\log2\sqrt{2A_{ex}/(M_sH_k)}.
\end{equation} 

Finally, the thermal stability is directly obtained as
\begin{equation}
\label{eq3}
    \Delta = \frac{\sigma_{dw}D\tau}{k_BT} = \frac{\sqrt{8M_sH_kA_{ex}}D\tau}{k_BT},
\end{equation}
where $H_k$ is the magnetic anisotropy, $A_{ex}$ is the exchange stiffness ($4.5~$pJ/m), and $D$ is the critical dimension of the pillar ($38.9~$nm). 
All the variables but the magnetic anisotropy $H_k$ are assumed based on those measured for our samples, for detailed measurements please see\,\cite{Choi2023}. 
To account for the impossibility to switch all the bits at large negative magnetic fields, we also divide equation \ref{eq1}  by a normalization constant. Under the above considerations, the anisotropy fields ($H_k$) obtained by fitting are  $\mu_0H_k= 440\pm2~$mT  for process~1 and $\mu_0H_k=533\pm2~$mT for process~2, which correspond to thermal stabilities ($\Delta$) of 49 and 54 respectively.  The domain wall widths ($\delta_W$) are also estimated to be 6 nm and 5 nm for process 1 and 2 respectively, a bit smaller than the estimated by Mihajlovic \textit{et al.}\cite{mihajlovic2020thermal}  for a slightly different film  stack ($\sim$12 nm).

To perform the simulations we used the  Magpylib library\,\cite{magpylib2020}. 
In these simulations the consecutive pillars are separated by 200 nm and each pillar is described with three cylinders with different thicknesses, diameters and magnetizations representing the FL, RL and HL. 
For the simulation shown in \figfive{}, we considered a thickness of $1.2$, $1.4$, and $3.8~$nm and a magnetization of $\pm1.175$, $+0.790$, and $+0.550~$MA/m for the FL (P or AP state), RL and HL respectively. 
For the critical dimension of the different pillars we assumed a normal distribution centered at $38.1~$nm with a standard deviation of $0.8~$nm which we base on SEM images similar to the one shown in \figtwo{b}.  
For FL and RL, we also considered a normal distribution in the magnetization direction centered at $0~^\circ$ (out of plane) with a standard deviation of $10~^\circ$ and $5~^\circ$ respectively. 
For the FL, a normal distribution in the absolute value of the magnetization centered at  $1.175~$MA/m  with a standard deviation of $0.235~$MA/m was also considered.  
Finally, to mimic the NV center, the magnetic field was measured at $151~$nm from the FL and at $54.5^\circ$ in respect to out of plane direction.
\begin{acknowledgments}
We would like to thank Tobias Sjölander, Siamak Salimy and Steven Lequex for useful discussions.The samples fabricated for this work were supplied by imec’s Industrial Affiliation Program on STT-MRAM devices. 
In addition,  the authors would like to acknowledge the support of imec’s fab, line, and hardware teams for the manufacture of these device arrays.
\end{acknowledgments}

\nocite{*}
\bibliography{paper-bib}% Produces the bibliography via BibTeX.
\end{document}